\documentclass[a4paper,10pt]{article}
\usepackage[utf8x]{inputenc}
\usepackage{amsmath}
\usepackage{graphicx}
\usepackage{graphics}
\usepackage{amssymb}\usepackage{graphicx}
\usepackage{tabularx}
\usepackage{tikz}
\usepackage{textpos}
\usepackage{bbold}
\usepackage[nodisplayskipstretch]{setspace}
\usepackage{pdflscape}
\usepackage[toc,page]{appendix}
\usepackage{authblk}
\newcommand{\ketbra}[2]{|{#1}\rangle \langle {#2}|}
\newcommand{\bra}[1]{\langle {#1}|}
\newcommand{\ket}[1]{| {#1} \rangle}
\newcommand{\tket}[2]{\ket{\widetilde{#1_{#2}}}}
\newcommand{\braket}[2]{\langle #1 | #2 \rangle}
\newcommand{\tbrakett}[4]{\langle \widetilde{#1_{#2}} | \widetilde{#3_{#4}} \rangle}
\newcommand{\me}{\mathrm{e}}
\newcommand{\pd}[1]{\frac{d #1 }{d  t}}
\title{Minimum Error Discrimination of Linearly Independent Pure States: Analytic Properties of POVM}
\author[1]{Tanmay Singal\thanks{stanmay@imsc.res.in}}
\author[1]{Sibasish Ghosh\thanks{sibasish@imsc.res.in}}
\affil[1]{Optics and Quantum Information Group, Institute of Mathematical Sciences, CIT Campus, Taramani, Chennai, 600 113, India}

\begin{document}

\maketitle

\begin{abstract}
The optimization conditions for minimum error discrimination of linearly independent pure states comprise of two kinds: stationary conditions over the space of rank one projective measurements and the global maximization conditions. A discrete number of projective measurments will solve th former of which a unique one will solve the latter. In the case of three real linearly independent pure states we show that the stationary conditions translate to a system of simultaneous polynomial (non linear) equations in three variabes thus explaining why it's so difficult to obtain a closed-form solution for the optimal POVM. Additionally, our method suggests that as an ensemble of LI pure states is varied as a smooth function of some independent parameters, the optimal POVM will also vary smoothly as a function of the same parameters. By employing the implicit functions theorem we exploit this fact to obtain a technique to find the solution of MED of LI pure states by dragging the solution from a known example (say, 
pure orthogonal states) to any general linearly indepenent ensemble of pure states in the same Hilbert space. By employing RK4 to solve the first order coupled non-linear differential equations find that the resulting error is within the RK4 error performance. 

\end{abstract}

\section{Introduction}
\label{intro}

Minimum Error Discrimination is one of the oldest problems in quantum state discrimination. The problem arises due to the vectorial nature of states which makes them indistinguishable through measurement. To infer what the state is, one has to perform measurement. Non-orthogonality of states implies that the measurement operators cannot perfectly distinguish one state from the other leading to errors in the discrimination procedure. Different measurement strategies have different performance strength (measured in terms of the average probability of error or the average probability of success). Given that the states cannot be distinguished perfectly there must be some measurement criterion which one could adopt to attain a maximum probability of success. To find what this measurement strategy is, is the problem of quantum state discrimination.

The setting in minimum error discrimination or quantum hypothesis testing or ambiguous state discrimination is the following: A has a fixed ensembled of states $\{ \rho_1, \rho_2, \cdots, \rho_m\}$ from which she selects one with certain probability $\{p_1, p_2, \cdots, p_m\}$ ($p_i\; >\;0,\; \sum_i \;p_i\; =\;1$)  respectvely and gives it to B. B knows that A has selected this state from that fixed set with those probabilities and his job now figure out which state he has been given using an m element POVM. There is a one-to-one correspondence between elements in A's ensemble $\{ p_i, \rho_i \}_{i=1}^{m}$ and B's POVM elements $\{ \Pi_i\}_{i=1}^{m}$ so that when the ith measurement outcome clicks, B infers A gave him the ith state from her ensemble. One can infer that the non-orthogonality of the ensemble elements implies that errors are likely to occur. B's job is to now find the optimal POVM for maximizing the average probability of success of his outcome.

There are other variants to the state discrimination problem\cite{Per}\cite{Croke}. The most popular alternative of them is called unambiguous state discrimination. The idea is that the measurement outcomes are now m+1 in number where, as in the MED case, there is a one-to-one correspondence between ensemble elements and the first m POVM elements. In this problem, the POVM is designed such that only when A sends B the ith state will the ith POVM element click, otherwise it won't. The trade-off is the m+1th element in the ensemble which is the "inconclusive" outcomer result i.e. B can say nothing about the state A sent him when this POVM element clicks. Heuristically one can expect that a set of linearly dependent states cannot be unambiguously discriminated in this manner, an idea that was proven true later\cite{Chef}. The more recent proposal of maxmimum confidence measurements is the application of unambiguous state discrimination to linearly dependent ensembles i.e. when linearly dependent then the measurement which maximizes the confidence of the state measured being the correct one.   

Coming back to MED, necessary and sufficient conditions which the optimal POVM has (or have) to satisfy were given by Holevo\cite{Hol} and Yuen\cite{Yuen} independently. The latter cast MED into a linear programming (and now SDP) problem for which numerical solutions can now be approximated within polynomial time. While there are quite a number of numerical techniques to obtain the optimal POVM upto very good approximation\cite{Eldar, Jezek, Hel2}, for very few ensembles has the MED problem been solved analytically. Compare that with the unambiguous state discrimination for which many more general ensembles have been solved.\cite{Per, Bergou, Ray, Herzog,Pang,Janos} Some of these include ensemble of two states\cite{Hel}, ensembles whose density matrix is maximally mixed state\cite{Yuen}, equiprobable ensembles that lie on the orbit of a unitary\cite{Ban},\cite{Chou},\cite{BS} and recently three mixed qubit states\cite{Ha}. For linearly independent pure state ensembles just the two-state ensemble problem 
has been solved but nobody has been able to give a solution for three linearly independent pure states.

We are motivated to understand what makes the three linearly independent pure state problem so difficult. For this purpose the problem is cast in different settings: (i) First it is cast in a geometric setting i.e. we make use of the geometry of the qutrit bloch sphere \cite{Q} to see what the problem looks like. Corresponding to the necessary and sufficient conditions given by Holevo and Yuen we get a set of simultaneous multivariate non-linear polynomial equations in terms of the components of the 8 dimensional qutrit bloch vectors of three states dual to the linearly independent pure states of the ensemble. It can be seen how complicated the problem here when contrasted with the simplicity of the two state discrimnation problem which is trivial once the problem is cast in the qubit state space.(ii) Corresponding to the generalized pretty good measurements \cite{Carlos}/ Belavkin weighted square root measurements \cite{Bela} one can obtain two set of real simultaneous multivariate 
polynomial equations - the more complicated one of which is essentially an attempt to find the inverse of the mapping $\widetilde{p}_i \longrightarrow  p_i$ given in \cite{Carlos}, while the other one is much simpler. In (i) and (ii) each set of real simultaneous multivariate polynomial equations have a discrete solution set of which a subset corresponds to m-element POVMs and of that a proper subset corresponds to the optimal POVM. The other POVMs have a physical interpretation: they are the stationary points of the objective function in the space of extremal m-element POVMs. When m = 2, there are only two possible solutions i.e. two such stationary points - one corresponding to the global minima and the other corresponding to the global maxima; and one is obtained from the other by interchanging indices of POVM elements of the latter. As soon as m=3, there can be multiple stationary points in the space of extremal POVMs some of which could be local maxima, local minima, saddle poitns, points of 
inflection, the global maxima and minimum etc. Thus all this information is encoded in equations. Given this nature of the solution set, it is unlikely that one can find a set of simpler equations to solve the problem. Add to this the problem that the set of reals is not algebraically closed i.e. real polynomial equations can have complex solutions. Thus often some solutions in the solution set are complex which then don't correspond to any POVM. 

One of the methods we used to obtain these equations suggests a way to analyze the analytic properties of the optimal POVM as a function of the ensemble i.e. to analyze how ensemble varies when the ensemble of LI pure states is varied. These ensembles are parameterized using their gram matrices. For studying the analytic behaviour, we employ the implicit function theorem which tells us that the optimal POVM will vary analytically as the ensemble is varied according to some parameterization. Using IFT we obtain a set of first order non-linear coupled differential equations using which we can "drag" the optimal POVM from some enseble to another. We employ RK4 to drag the optimal POVM from an equiprobable orthogonal ensemble of states to a general linearly independent ensemble. We note that the error increases slower than expected with the number of iterations telling us that this method is reliable to obtain MED for some LI pure ensemble upto desired accuracy. 

The paper is divided as follows: First we go into detail about what MED is. Then we give the optimizing conditions and specify what they look like for LI pure ensembles. Then we give a brief description on LI pure ensembles. We formulate the MED problem for linearly independent states using geometry of the qutrit state space and later in what we call the gram matrix method. In both cases we come across simultaneous real multivariate polynomial equations which have a zero dimensional variety but mutliple solutions. We note that while some of the solutions are complex and discardable the others correspond to some rank-one projective measurement and infer that all of them are stationary points of the objective function in the space of rank-one projective measurements. Moving on, we find a way to represent linearly independent ensembles using an equivalence class of trace one positive definite matrices and use that along with the implicit function theorem to show that the optimal POVM will vary analytically as a 
function of the ensemble. Then again using IFT we obtain a set of first order non-linear coupled differential equations which can be used to drag the solution from one ensemble to another. Starting with an equiprobable orthogonal ensemeble we employ RK4 to do this "dragging" and we note that we get a decent performance where the errors remain below expected. 

\section{Minimum Error Discrimination: The Problem}
\label{MEDP}
A has a device that prepares a quantum state, $\rho_i$ with probability $p_i$ from a fixed set, $\{ \rho_1, \rho_2,..., \rho_m \}$. Here $p_i > 0 \, ,\forall \, 1 \leq i \leq m$ and $\sum_{i=1}^{m} p_i=1$. Let supp$(\rho_i)= \mathcal{H}_i \,$, rank$(\rho_i)=r_i$. Also, let $\mathcal{H} =$ span$( \bigcup_{i=1}^{m} \mathcal{H}_i )$ and dim$(H)=n$.

A obtains a quantum state $\rho_i$ with probability $p_i$ from this device and gives this state to B without telling the latter which state he's being given. B knows that this state is from A's device and wants to know which state A has given him. For that A has to "distinguish" the state he's been given from the other ones in $\{p_i, \rho_i \}_{i=1}^{m}$ by performing a measurement on his state. This measurement is usually a generalized measurement i.e. a POVM. The measurement scheme is as follows: this measurement has m distinct outcomes each of which are indexed in such a way that the ith outcome indicates that he has been given $\rho_i$.

In case the states $\{ \rho_i \}_{i=1}^{m}$ are pairwise orthogonal i.e. $Tr( \rho_i \rho_j) =0, \, \forall \, 1 \leq i,j \leq m$, B can choose his POVM to be $\{ P_i \}_{i=1}^{m}$ where $P_i$ is a projector on the subspace $\mathcal{H}_i$ and $ P_i P_j = \delta_{i,j} P_i, \, \forall \, 1 \leq i,j \leq m$. In such a case $Tr(P_i \rho_j)=\delta_{i,j}$ leading to \emph{always} correct inference and we say that elements from the ensemble can be distinguished perfectly.

In case the states $\{\rho_i \}_{i=1}^m$ aren't pairwise orthogonal (as is usually the case) there is no measurement that can distinguish perfectly between them. This means that it may so happen that despite being given $\rho_i$, B's measurement output is j, leading to an error. 

The average probability of error is given by:

\begin{equation}
\label{Pe}
P_e = \sum_{\substack{i,j=1 \\ i\neq j}}^m p_i Tr( \rho_i \Pi_j)
\end{equation}

where $\{ \Pi_i \}_{i=1}^{m}$ represents an m element POVM with $ \Pi_i \geq 0$ and $ \sum_{i=1}^{m} \Pi_i = \mathbb{1}$. 

The average probability of success is given by:

\begin{equation}
 \label{Ps}
P_s =  \sum_{i=1}^m p_i Tr( \rho_i \Pi_i)
\end{equation}

Both probabilities sum up to 1:

\begin{equation}
\label{Sum1}
P_s + P_e =1
\end{equation}

It is worth noting that $P_s$ (and $P_e$) is a linear function of the m-element POVM $\{ \Pi_i \}_{i=1}^{m}$ and that the set of m-element POVMs form a compact convex set. From \eqref{Sum1} it immediately follows that maximization of $P_s$ over these m-element POVMs implies the minimization of $P_e$ over the same. 

\begin{equation}
\label{Pmax}
P_{s}^{\text{max}} = \text{Max}_{\substack{ \{ \Pi_i \}_{i=1}^{m} \\ \Pi_i \geq 0 \\ \sum_{i} \Pi_i = \mathbb{1}} } P_s = 1- P_{e}^{\text{min}}
\end{equation}

Thus B's task is finding the optimizing m-element POVM. Given that the set of m-element POVMs forms a compact set, we know that such an m-element POVM must exist. 

\section{The Optimum Conditions}
\label{OPTC}
Thus B's task is an optimization problem \eqref{Pmax} over a constrained set $\Pi_i \geq 0, \: \: \sum_{i} \Pi_i = \mathbb{1}$. To every constrained optimization problem (called the primal problem) there exists a corresponding dual problem which provides a lower bound (if primal problem is a constrained minimization) or an upper bound (if the primal problem is a constrained maximization) to the quantity being optimized in the primal problem. Under certain conditions these bounds are tight implying that one can obtain solution for the primal problem from its dual. We then say that there is no duality gap between both problems. 

For MED there is no duality gap and the dual problem can be solved to obtain optimal POVM. This dual problem is given as follows \cite{Yuen}: 
\begin{equation}
\label{dual}
\text{Min} \: \text{Tr}(Z) \: \ni \; Z \geq p_i \rho_i \, \forall\, 1 \leq i \leq m 
\end{equation}

Also the optimal m-element POVM will satisfy the complementarity slackness condition:

\begin{equation}
\label{cslack}
(Z- p_i \rho_i)\Pi_i= \Pi_i(Z-p_i \rho_i)=0, \, \forall \, 1\leq i \leq m
\end{equation}

Now summing over $i$ in \eqref{cslack} and using the fact that $ \sum_{i=1}^{m} \Pi_i = \mathbb{1}$ we get:

\begin{equation}
\label{Z}
Z= \sum_{i=1}^{m} p_i \rho_i \Pi_i = \sum_{i}^{m} \Pi_i p_i \rho_i
\end{equation}

From \eqref{cslack} we get

\begin{align}
& \Pi_j ( Z - p_i \rho_i) \Pi_i = \Pi_j ( Z - p_j \rho_j) \Pi_i \notag \\
\label{St}
& \Rightarrow \Pi_j ( p_j \rho_j - p_i \rho_i ) \Pi_i =0 , \forall 1 \leq i,j \leq m
\end{align}

\eqref{St} was derived by Holevo, separately, without using the dual optimization problem. \eqref{cslack} and \eqref{St} are equivalent to each other. These are necessary but not sufficient conditions. The solution set contains a finite number of m-POVMs of which a proper subset is optimal. This optimal POVM will satisfy the global maxima conditions given below:

\begin{align}
 & Z \geq p_i \rho_i \notag \\
\label{Glb}
 & \Rightarrow \sum_{k=1}^{m} p_k \rho_k \Pi_k - p_i \rho_i \geq 0 , \forall 1 \leq i \leq m     
\end{align}
 
Thus the necessary and sufficient conditions for the m-element POVM to globally maximize $P_s$ is are given by \eqref{St} and \eqref{Glb}. 

\subsection{Linearly Independent Pure State Ensembles}
\label{LIPSE}
Whatever has been said till now applies to the minimum error discrimination for any ensemble of states $\{ p_i > 0,\, \rho_i \}_{i=1}^{m}$. From here on we wish to examine minimum error discrimination of linearly independent pure states i.e. the ensemble is $ \{ p_i, \ketbra{\psi_i}{\psi_i} \}_{i=1}^{m}$ where $ \{ \ket{\psi_i} \}_{i=1}^{m}$ is a linearly independent set. This implies that $r_i = 1 \; \forall \; 1 \leq i \leq m$ and that $n=m$. Hence the space is now m-dimensional.

Continuous optimization problems\footnote{constrained or otherwise} generally yield two differnt kinds of conditions both of which need to be solved: the first one correspond to stationary conditions of the objective function in the space of feasible solutions\footnote{ The space of feasible solutions is the space which statisfies the constraints of the problem. In our case these constraints are $\Pi_i \geq 0$ and $\sum_{i} \Pi_i = \mathbb{1}$ i.e. the space of m element POVMs.}. These conditions are in the form of equations. The second correspond to the minimization or maximization of the objective function. These conditions are in the form of inequalities. One could adopt two different approaches for solving the problem at hand: (i) Solve both the stationary and the extrema conditions simultaneously. (ii) First solve the stationary conditions. Typically, the solution set of the stationary conditions will be discrete. Having obtained this solution set, test each solution individually 
for the maxima/ minima conditions.

Let's interpret what the equations \eqref{St} and \eqref{Glb} mean when the ensemble is that of linearly independent pure states.

It can be shown \cite{Carlos, Hel, Ken} that in this context,  \eqref{St} imply that the elements of the POVM have to be rank-one. Thus the POVM comprises of m rank-one elements and covers a space that is m-dimensional. This implies that the POVM has to be a rank-one projective measurement. Thus we can restrict the candidate set to the space of rank-one projective measurements. What does \eqref{St} imply further? It can be easily shown that \eqref{St} are conditions for the stationary points of the objective function $P_s$ in space of rank-one projective measurments. This describes fully the role played by \eqref{St} in the optimization problem. 

Once \eqref{St} has been satisfied by some rank-one projective measurements, \eqref{Glb} is the condition for the global maximum of $P_s$ on them. It has to be noted that this is different from the local maximum condition which wouldn't be sufficient for MED.

Solving optimization problems by (i) is more difficult than (ii), even numerically. Here we attempt to solve it analytically for the case where $n=3$ and ensemble is real.\footnote{A "real ensemble" implies that there exists a basis in which all states of the ensemble become real.}${ }^{,}$\footnote{Arguments in \cite{Carlos},\cite{Bela} imply that when dealing with a real ensemble the conditions, equations and solutions to the problem are entirely real as well. This doesn't serve any physical purpose; just is a simpler to solve.} We try solving the problem through both (i) and (ii). 

(i.a) The MED of many general families of ensembles of qubit states can be solved using the qubit-state space geometry \cite{Bae}\cite{Ha}\cite{Sam} \footnote{In fact for m=2, solution is trivial}. We investigate what the problem looks like for MED for linearly independent pure states in the qutrit state space. As can be expected, the richness of the qutrit space geometry makes the problem a lot more difficult. The advantages that the qubit space geometry has to offer, owing to its simplicity and ease of visualization, are lost when it comes to the qutrit space. \eqref{St} and \eqref{Glb} give us some polynomial equations and inequalities. Given these equations and inequalities one proceeds by first solving the equations and then testing the inequalities with each element in the solution set. Thus while we started with the intention of solving by (i) we ultimately have to resort to solve by (ii). One can formulate the problem in other means as we will describe below. Those equations are much simpler 
to solve. The reason why we made these equations explicit is to give the general picture of how complicated the problem can become for even simplest systems.

(i.b) Staring from \eqref{St} and \eqref{Glb}, we reformulate the problem adopting a representation in a special basis. Here we encounter some simultaneous non-linear equations. These equations aren't polynomial equations but can be manipulated to obtain polynomial equations. Yet one loses certain information through these manipulations\footnote{This manipulation involves squaring or cubing both sides of the equation resulting in the loss of some information.}.Some of the information lost pertains to the global maxima conditions \eqref{Glb}. Hence the new polynomial equation we obtain give us solutions for the stationary points among which one uniquely will correspond to the global maxima. Since the non-polynomial equations are more tedious to solve, it is better to opt for the polynomial equations. Once again, despite starting out to solve through (i), we resort to solving through (ii). 

(ii)The polynomial equations from (i) are very complicated. One desires much more simplified equations. The representation just used implies that one can obtain close cousin of equations to those obtained in (i.b) which are much simpler to solve. For real systems \eqref{St} are cast into a set of 3 simultaneous real trivariate polynomial equations with a discrete solution set\footnote{For complex systems, the number of variables and equations are six.}. Given that $\mathbb{R}$ isn't an algebraically closed field, this discrete solution set may include some complex solutions which are discarded. Each one of the remaining solutions is real and corresponds to a rank-one projective measurment where $P_s$ is stationary\footnote{i.e. $P_s$ is stationary in the space of rank-one projective measurements. On the other hand, $P_s$ cannot be stationary in the space of m-POVMs. This is because $P_s$ is linear in the m-POVMs and space of m-POVMs is convex. Had $P_s$ been stationary at any point in the m-POVM space, it 
would have to be constant for all points in the space. We know that this isn'true.}. Of these stationary points, a unique one is the global maximum whereas the others correspond to saddle points, points of inflection, local maxima, local minima and the global minimum. Each one of these solutions has to be individually tested for \eqref{Glb}. Obtaining solutions for these polynomial equations implies solving them symbolically. As we will see later, this is a very difficult task to accomplish\footnote{It needs to be mentioned that we obtained other polynomial equations for solving \eqref{St} too. The reason we avoid mentioning them explicitly is because they involve a greater number of real equations in real variables. Suffice it to say that whatever holds for \eqref{St} holds for any such polynomial set.}.

\subsection{Geometric Method}
\label{SSGM}
The geometry of the generalized Block sphere $\Omega_3$ for qutrits was completely specified in \cite{Q}. Here we go through it briefly:

Any density matrix $\rho$ has a unique representation:

\begin{equation}
\label{denqu}
\rho = \frac{1}{3} (\mathbb{1} + \sqrt{3} \vec{n}.\vec{\lambda})
\end{equation}

where $\lambda_i$ represent the Gell-Mann matrices and $\vec{n}$ represent a vector from $\mathbb{R}^8$ which satisfy the following properties:

\begin{align}
\label{qucon1}
\vec{n}.\vec{n} \leq 1 \\
\label{qucon2}
3 \vec{n}.\vec{n} - 2 (\vec{n}*\vec{n}).\vec{n} \leq 1 
\end{align}

\eqref{qucon1} is the condition for $\rho$ to be non-negative whereas \eqref{qucon2}\footnote{ $(\vec{n}*\vec{n})_{l}= \sqrt{3} d_{jkl} n_j n_k$ where $d_{jkl} = \frac{1}{4} Tr( \lambda_j \{ \lambda_k , \lambda_l \})$ } specifies the boundary and interior of $\Omega_3$. 

Let $\ketbra{\psi_i}{\psi_i}$ correspond to bloch vector ${\vec{n}}^{(i)}$ which saturates both \eqref{qucon1} and \eqref{qucon2}. For $ \{ \vec{n}^{(i)} \}_{i=1}^{3}$ to correspond to a set of linearly independent vectors, we require that $\rho'= \sum p_i \ketbra{\psi_i}{\psi_i}$ lies strictly in the interior of $\Omega$ i.e. let $\vec{n} = \sum_{i=1}^{3} p_i \, \vec{n}^{(i)}$, then $3 \vec{n}.\vec{n} - 2 (\vec{n}*\vec{n}).\vec{n} < 1 $. Here we don't impose the condition that $\ketbra{\psi_i}{\psi_i}$ be real. 

From \eqref{cslack} and \eqref{Glb} we know that the operators $\widetilde{\sigma}_i := Z - p_i \ketbra{\psi_i}{\psi_i}$ are positive semidefinite with kernel spanned by one dimensional Supp$(\Pi_i)$. We rewrite $\widetilde{\sigma}:= \kappa_i \sigma_i$ where $\text{Tr}(\sigma_i)=1$. Hence $\sigma_i$ is a rank two density operator. Since $\Pi_i$ are rank 1 they are extreme points on the boundary of the bloch sphere. Let bloch vectors corresponding to $\sigma_i$ be ${\vec{s}}^{(i)}$ and for $\Pi_i$ be  ${\vec{t}}^{(i)}$. It follows that 

\begin{equation}
\label{sb}
(3 {\vec{s}}^{(i)} -2 {\vec{s}}^{(i)}*{\vec{s}}^{(i)}).{\vec{s}}^{(i)}=1, \quad  {\vec{s}}^{(i)}.{\vec{s}}^{(i)} <1
\end{equation}
\begin{equation}
\label{tb}
(3 {\vec{t}}^{(i)} -2 {\vec{t}}^{(i)}*{\vec{t}}^{(i)}).{\vec{t}}^{(i)}=1,\quad  {\vec{t}}^{(i)}.{\vec{t}}^{(i)} =1
\end{equation}

Consider
\begin{equation}
\label{Zed}
Z = \frac{k_0}{3} (\mathbb{1} + \sqrt{3} \vec{k}.\vec{\lambda})
\end{equation}

where \begin{align}
& \text{Max }(p_1,p_2,p_3)\leq k_0  \leq 1  \notag \\
& \vec{k}.\vec{k} <1 \notag \\
\label{ink}
&(3 \vec{k}-2\vec{k}*\vec{k}).\vec{k}\leq 1       
      \end{align}
And hence we get

\begin{align}
\kappa_i= k_0-p_i \notag \\
\label{sk}
{\vec{s}}^{(i)}=\frac{k_0 \vec{k}-p_i {\vec{n}}^{(i)}}{\kappa_i}
\end{align}

Now ${\vec{t}}^{(i)}$ is a function of ${\vec{s}}^{(i)}$. Since ${\vec{s}}^{(i)}$ satisfies \eqref{sb}  ${\vec{t}}^{(i)}$ has to satisfy the following equation linear in its components:

\begin{equation}
\label{ot}
{\vec{s}}^{(i)}+{\vec{t}}^{(i)} + {\vec{t}}^{(i)}*{\vec{s}}^{(i)}=\vec{0}
\end{equation}

\eqref{sb} guarantees that there is a unique solution of \eqref{ot} for ${\vec{t}}^{(i)}$ and this solution automatically obeys \eqref{tb}. That said this functional dependence of ${\vec{t}}^{(i)}$ on ${\vec{s}}^{(i)}$ is extremely complicated. Now
\begin{equation}
\label{Pist}                                                                                                                                                                                                                                                                     
\sum_i \Pi_i = \mathbb{1} \; \Rightarrow \sum_i {\vec{t}}^{(i)} = \vec{0}                                                                                                                                                                                                                                                                      \end{equation}

The polytope formed by the points $\{ p_i {\vec{n}}^{(i)} \}_{i=1}^{3}$ and $\{ \kappa_i {\vec{s}}^{(i)} \}_{i=1}^{3}$ in $\mathbb{R}^8$ are congruent. In fact, the latter is a displaced mirror image of the former. 

\begin{align}
\label{xyz}
& p_i - p_j = \kappa_j -\kappa_i \\
\label{geo}
& p_i {\vec{n}}^{(i)}  - p_j {\vec{n}}^{(j)} = \kappa_j {\vec{s}}^{(j)} -\kappa_i {\vec{s}}^{(i)}
\end{align}

We also require that the identity should be resolved by the states orthogonal to $\sigma_i$ (i.e. $\Pi_i$). One of the main hurdles faced in the qutrit case is to \textbf{visualize} where ${\vec{t}}^{(i)}$ should be located so that $\text{Tr}(\sigma_i \Pi_i)=0$ is satisfied. Also, one needs to satisfy \eqref{Pist}. In the qubit case $\sigma_i$ would have been a rank-one pure state impling that ${\vec{s}}^{(i)}$ would be on the surface of the bloch sphere and ${\vec{t}}^{(i)}$ would be the former's corresponding anti-podal point. These simplicities are no more in the qutrit case. In general there are multiple set of states $ \{ \sigma_i \}_{i=1}^{3}$ for which some $\kappa_i$ can be found so that \eqref{xyz} and \eqref{geo} are satisfied. The problem now is to ensure that \eqref{Pist} too is satisfied. The only way one can proceed is to solve the the system of $9$ unknowns $k_0, \vec{k}$ for \eqref{ink},\eqref{sb},\eqref{ot} and \eqref{Pist}, where ${\vec{s}}^{(i)}$ is given by \eqref{sk}, algebraically. \
eqref{sb},\eqref{ot} and \eqref{Pist} are all polynomial equations whereas \eqref{ink} are inequalities. Satisfying these equations gives us the optimal POVM ${\vec{t}}^{(i)} \longrightarrow \Pi_i$. Explicit form of ${\vec{t}}^{(i)}$ in terms of components of ${\vec{s}}^{(i)}$ make the polynomial equations really tedious. It's desirable to obtain a much simpler set of equations. With that in mind we turn to a new method.

\subsection{Gram Matrix Method}
\label{SSGMM}
We first talk about the general n-dimensional systems. Later on we specialize for n=3.

We wish to obtain the optimal POVM (which is a rank-one projective measurement) for MED for an ensemble $\{p_i, \ketbra{\psi_i}{\psi_i} \}_{i=1}^{m}$ where $\{ \ket{\psi_i} \}_{i=1}^{m}$ is a linearly independent set. Let $ \tket{\psi}{i} = \sqrt{p_i} \ket{\psi_i} , \; \forall \; 1\leq i \leq m$. Since $\{ \tket{\psi}{i} \}_{i=1}^{m}$ form a linearly independent set, there exists a corresponding unique ordered set $\{ \tket{u}{i} \}_{i=1}^{m}$ where $  \tket{u}{i} \in \mathcal{H} $ such that

\begin{equation}
\label{orth}
\tbrakett{\psi}{i}{u}{j}= \delta_{i,j} \;, \; \forall \; 1 \leq i,j \leq m
\end{equation}
Let G denote the gram matrix of $\{ \tket{\psi}{i} \}_{i=1}^{m}$.The matrix elements of G are hence given by
\begin{equation}
\label{Gram}
\text{G}_{ij} = \tbrakett{\psi}{i}{\psi}{j} \;, \; \forall \; 1 \leq i,j \leq m
\end{equation}
Since $\{ \tket{\psi}{i} \}_{i=1}^{m}$ is linearly independent set, $G \geq 0$. The gram matrix of $\{ \tket{u}{i} \}_{i=1}^{m}$ is $\text{G}^{-1}$. Any ordered orthonormal basis $\{ \ket{v_i} \}_{i=1}^{m}$ can be represented as follows:

\begin{equation}
\label{v}
\ket{v_i} = \sum_{\substack{j=1}}^{m} (\text{G}^{\frac{1}{2}} U)_{ji} \tket{u}{j}
\end{equation}

where $\text{G}^{\frac{1}{2}}$ is the positive square root of G and U is an n dimensional unitary matrix. U captures the unitary degree of freedom of the ordered orthonormal basis. One can easily check that $\braket{v_i}{v_j}=\delta_{i,j}$. Any such an ordered orthonormal basis corresponds to an n-element rank-one projective measurement:

\begin{equation}
\label{btp}
\{ \ket{v_i} \}_{i=1}^{m} \longrightarrow  \{ \ketbra{v_{i}}{v_{i}} \}_{i=1}^{m}  
\end{equation}

In particular, when $U=\mathbb{1}$ we obtain the pretty good measurment associated with $\{ \tket{\psi}{i} \}_{i=1}^{m}$. Now consider that one can change the phase factors of the ONB vectors by appending a diagonal unitary, $U'$ on the right of U in \eqref{v} where $U'_{jk} = \delta_{j,k} \me^{ i \theta_j}$. This implies  $\{ \ket{v_j} \}_{j=1}^{m} \longrightarrow \{ \me^{i \theta_j} \ket{v_j} \}_{j=1}^{m}$. Changing these phase factors doesn't change the rank-one projective measurement the basis corresponds to. If one were to use this rank-one projective measurment for MED, the probability of success on average is given by:
\begin{equation}
P_s = \sum_{\substack{i=1}}^{n} | \braket{ \widetilde{ \psi_i}}{v_i}|^{2} = \sum_{\substack{i=1}}^{n} | (\text{G}^\frac{1}{2} U)_{ii}|^2
\end{equation}
The terms $ | (\text{G}^\frac{1}{2} U)_{ij}|^2$ where $i\neq j$ is the error probabillity that B's measurement yields j despite A sending him $\rho_i$. Note that when U is unity, the probability of ith POVM-element clicking while $\rho_j$ was sent is equal to the probability of jth POVM-element clicking while $\rho_i$ was sent.

Now using \eqref{v} in \eqref{St} we get:
\begin{equation}
\label{StG}
(\text{G}^{\frac{1}{2}} U)_{jj} (\text{G}^{\frac{1}{2}} U)_{jk}^{*}=(\text{G}^{\frac{1}{2}} U)_{kj}(\text{G}^{\frac{1}{2}} U)_{kk}^{*} \; \forall \; 1 \leq j,k \leq m
\end{equation}
which is a condition on $U$.

For the optimal POVM $\{ \ketbra{\widetilde{v}_{i}}{\widetilde{v}_{i}} \}_{i=1}^{m}$, Carlos Mochon \cite{Carlos} showed that
\begin{equation}
\label{diag0}
| \braket{\widetilde{\psi_i}}{\widetilde{v}_{i}}|^2 > 0 \; \forall \; 1 \leq i \leq m
\end{equation}
 
Suppose $\widetilde{U}$ corresponds to the optimal rank one projective measurment through the correspondence given by \eqref{btp}. Then \eqref{diag0} implies that the diagonal elements of  $\text{G}^{\frac{1}{2}} \widetilde{U}$ have to be non-zero. Also one can use $U'$ to make these diagonal elements positive. Thus \eqref{StG} becomes:

\begin{equation}
\label{StG1}
(\text{G}^{\frac{1}{2}} \widetilde{U})_{j,j} (\text{G}^{\frac{1}{2}} \widetilde{U})_{j,k}^{*}=(\text{G}^{\frac{1}{2}} \widetilde{U})_{k,j}(\text{G}^{\frac{1}{2}} \widetilde{U})_{k,k} \; \forall \; 1 \leq j,k \leq m
\end{equation}

Now we have the matrix equation:

\begin{equation}
\label{A1}
(\widetilde{U}^\dag \text{G}^{\frac{1}{2}}) G^{-1} (\text{G}^{\frac{1}{2}} \widetilde{U}) = \mathbb{1}
\end{equation}

Let D be a postive diagonal matrix comprising the diagonals of $\text{G}^\frac{1}{2} \widetilde{U}$:

\begin{equation}
\label{D}
\text{D}_{i,j} = \delta_{i,j} (\text{G}^{\frac{1}{2}} \widetilde{U})_{i,j} \; \forall \; 1 \leq i,j \leq m
\end{equation}

From \eqref{StG1} we infer that the matrix $\text{DG}^{\frac{1}{2}} \widetilde{U}$ is hermitian. Also since $\text{D} >0$, $\text{D}^{-1}$ exists. From \eqref{A1} we then get:

\begin{equation}
\label{B1}
(\widetilde{U}^\dag \text{G}^{\frac{1}{2}}\text{D}) (D G D)^{-1} (\text{D G}^{\frac{1}{2}} \widetilde{U}) = \mathbb{1}
\end{equation}

From \eqref{B1} we can see that $\text{DG}^{\frac{1}{2}}\widetilde{U}$ is a hermitian square root of the matrix $\text{DGD}$. If we represent $ D_{ii} = a_i \; \forall 1 \leq i \leq m$ we see that the diagonal elements of $\text{DG}^{\frac{1}{2}}\widetilde{U}$ are $a_{i}^{2}$. Thus we can solve the stationary condition \eqref{StG1} by finding a set of positive real numbers $\{ a_i \}_{i=1}^{m}$ such that a hermitian square root of the matrix $\text{DGD}$ has diagonal entries equal to $ a_{i}^{2}$. Having obtained such a set of positive numbers, one can proceed to find the associated rank-one projective measurment which will satisfy the stationary conditions \eqref{St}. 

For a given G there are many such sets of positive reals $\{ a_i \}_{i=1}^{m}$ which satisfy this desired property. Each of these different sets correspond to different rank-one projective measurements each of which will satisfy the stationary condition \eqref{St}. In keeping with the fact that only one POVM can be optimal, only one of these rank-one projective measurements satisfy the global maxima condition \eqref{Glb}. Carlos Mochon and Belavkin \cite{Carlos}\cite{Bela} noted that this optimal solution corresponds to the case where $\text{DG}^{\frac{1}{2}}\widetilde{U}$ is the positive square root of $\text{DGD}$. The corresponding projective measurment is the pretty good measurment for another ensemble with the same states $\{ \ket{\psi_i} \}_{i=1}^{m}$ but with different probabilities. 

\subsubsection{}
\label{SSGMM1}
Here we obtain algebraic equations to solve for the set $\{ a_1, a_2, a_3 \}$ for the case $m=3$ and where the ensemblse is real. G is given to us. $ a_1, a_2, a_3$ are our unknowns. The equations are obtained as follows:
\begin{equation}
\label{Gme}
(\sqrt{\text{DGD}})_{ii}= a_{i}^{2} \; \forall \; 1 \leq i \leq 3
\end{equation}

The expressions for $(\sqrt{\text{DGD}})_{ii}$ in terms of the matrix elements of G and in terms of the unknowns $a_1, a_2, a_3$ are extremely complicated. Here we do not write the full expression down, but symbolically denote matrix elements of $(\sqrt{\text{DGD}})_{ii}$ as:

\begin{equation}
\label{ex}
(\sqrt{\text{DGD}})_{ii} = \sum_{j}^{3} \sqrt{ \lambda^{(j)}} |{\zeta^{(j)}}_{i}|^{2}
\end{equation}

where $\lambda^{(j)}$ are the eigenvectors of DGD, $\vec{\zeta^{(j)}}$ the corresponding eigenvector and ${\zeta^{(j)}}_i$ the ith component of this eigenvector. Now $\lambda^{(j)}$ and ${\zeta^{(j)}}_i$ are complicated functions of the unknowns $a_1, a_2 \text{ and } a_3$ and matrix elements of G.\footnote{In \cite{Carlos} an invertible mapping is made from and to the space of non-zero m probabilities in the form: $p_i = C \frac{\widetilde{p_i}}{{({\widetilde{G}}^{\frac{1}{2}})}_{ii}}$ where $1 \leq i \leq m$. Then \eqref{ex} is precisely the exercise of obtaining the inverse function.} The equations \eqref{Gme} subsume not only \eqref{St} but the global maxima conditions \eqref{Glb}.

Given the complicated expression of the equations in $a_1, a_2, a_3$, one is discouraged from solving these simultaneous equations analytically. It would be better to undertake the tedious exercise to cast them into polynomial equations by rearranging terms in the equations and squaring and cubing the equations. This necessarily means that the information of  $(\sqrt{\text{DGD}})_{ii}$ corresponding to matrix elements of the \textbf{positive} square root of DGD is lost. Thus solutions from the resulting polynomial equations will always solve \eqref{St} but only one of them will solve \eqref{Glb}. 

\subsubsection{}
\label{SSGMM2}
In this subsection we obtain a different set of simultaneous multivariate polynomial equations. These are more easily obtained. Just that the number of unknowns increases to more than m. To reduce the number of unknowns we restrict ourselves to the case where $m=3$ and where the gram matrix G is real.

From \eqref{StG1} we infer that the matrix $\text{G}^{\frac{1}{2}} \widetilde{U} \text{D}^{-1}$ is hermitian with unit diagonal elements. 

From  \eqref{A1} we get
\begin{equation}
\label{A2}
(\text{D}^{-1}\widetilde{U}^\dag \text{G}^{\frac{1}{2}}) \text{G}^{-1}(\text{G}^{\frac{1}{2}} \widetilde{U}\text{D}^{-1}) = \text{D}^{-2}
\end{equation}

G being real implies that $\text{G}^{-1}$ is also real. In that case the matrix $\text{G}^{\frac{1}{2}} \widetilde{U} \text{D}^{-1}$ must be real symmetric with unit diagonal elements. 

Let \begin{equation}
\label{MD}
\text{G}^{\frac{1}{2}} \widetilde{U} \text{D}^{-1} =
 \begin{pmatrix}
1 & \alpha & \beta \\
\alpha & 1 & \gamma \\
\beta & \gamma & 1
\end{pmatrix}
\end{equation}
where $\alpha$, $\beta$ and $\gamma$ are real numbers. Then \eqref{A2} becomes:

\begin{align}
\label{A3}
& \begin{pmatrix}
1 & \alpha & \beta \\
\alpha & 1 & \gamma \\
\beta & \gamma & 1
\end{pmatrix}
 \begin{pmatrix}
(\text{G}^{-1})_{11} & (\text{G}^{-1})_{12} & (\text{G}^{-1})_{13} \\
(\text{G}^{-1})_{21} & (\text{G}^{-1})_{22} & (\text{G}^{-1})_{23} \\
(\text{G}^{-1})_{31} & (\text{G}^{-1})_{32} & (\text{G}^{-1})_{33}
\end{pmatrix}
 \begin{pmatrix}
1 & \alpha & \beta \\
\alpha & 1 & \gamma \\
\beta & \gamma & 1
\end{pmatrix}  \notag \\
& =  \begin{pmatrix}
(D_{11})^{-2} & 0 & 0 \\
0 & (D_{22})^{-2} & 0 \\
0 & 0 & (D_{33})^{-2}
\end{pmatrix} 
\end{align}

Here we get $6$ different equations:

\begin{align}
\alpha^2 (\text{G}^{-1})_{12} \; + \; \alpha( (\text{G}^{-1})_{11} \; + \;(\text{G}^{-1})_{22} \; + \; (\text{G}^{-1})_{13}\beta \;+ \; (\text{G}^{-1})_{23} \gamma) \notag \\
 \label{E1}
\; + \; (\text{G}^{-1})_{33} \beta \gamma \; + \; (\text{G}^{-1})_{23} \beta \; + \; (\text{G}^{-1})_{13} \gamma \; + \; (\text{G}^{-1})_{12} =0 \\
\beta^2 (\text{G}^{-1})_{13} \; + \; \beta( (\text{G}^{-1})_{11} \; + \;(\text{G}^{-1})_{33} \; + \; (\text{G}^{-1})_{23}\gamma \;+ \; 
(\text{G}^{-1})_{12} \alpha) \notag \\
 \label{E2}
\; + \; (\text{G}^{-1})_{22} \alpha \gamma \; + \; (\text{G}^{-1})_{12} \gamma \; + \; (\text{G}^{-1})_{23} \alpha \; + \; (\text{G}^{-1})_{13} =0 \\
\gamma^2 (\text{G}^{-1})_{23} \; + \; \gamma( (\text{G}^{-1})_{22} \; + \;(\text{G}^{-1})_{33} \; + \; (\text{G}^{-1})_{13}\beta \;+ \; (\text{G}^{-1})_{12} \alpha) \notag \\
 \label{E3}
\; + \; (\text{G}^{-1})_{11} \alpha \beta \; + \; (\text{G}^{-1})_{12} \beta \; + \; (\text{G}^{-1})_{13} \alpha \; + \; (\text{G}^{-1})_{23} =0\\
(\text{G}^{-1})_{22} \alpha^2 \; + \; (\text{G}^{-1})_{33} \beta^{2} + 2 \alpha \beta (\text{G}^{-1})_{23} \; + \; 2 \alpha (\text{G}^{-1})_{12} \; + \; 2 \beta (\text{G}^{-1})_{13} \notag \\
\label{E4}
 \; + \; (\text{G}^{-1})_{11} = (\text{D}_{11}^{-2}) \\
(\text{G}^{-1})_{11} \alpha^2 \; + \; (\text{G}^{-1})_{33} \gamma^{2} + 2 \alpha \gamma (\text{G}^{-1})_{13} \; + \; 2 \alpha (\text{G}^{-1})_{12} \; + \; 2 \gamma (\text{G}^{-1})_{23} \notag \\
\label{E5}
 \; + \; (\text{G}^{-1})_{22} = (\text{D}_{22}^{-2}) \\
(\text{G}^{-1})_{11} \beta^2 \; + \; (\text{G}^{-1})_{22} \gamma^{2} + 2 \beta \gamma (\text{G}^{-1})_{12} \; + \; 2 \beta (\text{G}^{-1})_{13} \; + \; 2 \gamma (\text{G}^{-1})_{23} \notag \\
\label{E6}
 \; + \; (\text{G}^{-1})_{33} = (\text{D}_{33}^{-2})
\end{align}

We obtain solutions for $(\alpha, \beta, \gamma)$ from \eqref{E1}, \eqref{E2} and \eqref{E3} and then substitute them in the nex three equations \eqref{E4}, \eqref{E5} and \eqref{E6} to obtain the values of $\text{D}_{ii}$ from which we can now find out what $\text{G}^{\frac{1}{2}} \widetilde{U}$ is and from there obtain the optimal projective measurement.

One obtains 8 solutions for the equations  \eqref{E1}, \eqref{E2} and \eqref{E3}  for  $(\alpha, \beta, \gamma)$. In certain instances some of the solutions in the solution set aren't real. By substituting these complex $(\alpha, \beta, \gamma)$ in \eqref{MD} we obtain a complex symmetrix matrix which isn't hermitian as was required. Hence such complex solutions are unphysical in the sense that they don't correspond to some rank-one projective measurement. All the remaining solutions are real and correspond to some rank-one projective measurment. The uniqueness of the optimal POVM for linearly independent pure state ensembles implies that only one of these real solutions corresponds to the optimal POVM. The physical significance of the other real solutions is that they provide those points in the space of rank-one projective measurements which are stationary for the function $P_s$ as given by \eqref{Ps}. How does one identify the solution corresponding to the optimal POVM? From the previous section we know 
that the matrix $\text{DG}^{\frac{1}{2}} \widetilde{U}$ has to be a positive matrix to correspond to the optimal POVM. Since $\text{DG}^{\frac{1}{2}} \widetilde{U}$ and $\text{G}^{\frac{1}{2}} \widetilde{U} \text{D}^{-1}$ are related by a congruence transformation, the optimal solution in our case must correspond to those values of $( \alpha, \beta, \gamma )$ for which the matrix $\text{G}^{\frac{1}{2}} \widetilde{U} \text{D}^{-1}$ is positive definite. The uniqueness of the optimal POVM then implies that there's only one solution of $( \alpha, \beta, \gamma )$ for which the matrix $\text{G}^{\frac{1}{2}} \widetilde{U} \text{D}^{-1}$  is positive definite.

Symbolically the reduced Groebner basis of the ideal of these polynomials contains 14 different elements. "Solving" the system would imply that one has to divide the parameter space into disjoint "cells" and specify a regular chain for each such cell\cite{Chen}. In doing even this, one cannot be sure if the resulting polynomials will be of degree $\leq$ 4 in the main variable in each polynomial (with respect to some suitable monomial ordering criteria). Hence we see that even by restricting ourselves to $n=3$ and a real gram matrix, obtaining a closed form solution for the optimal POVM is still very tough. 

\section{Analytic Properites of Optimal POVM for Pure State Ensembles}
\label{Anal}
In this section we analyze the optimal m-POVM for MED of m linearly independent pure states as a function of the ensemble. To do that we first describe this space of m-element ensembles. We denote it by $\mathfrak{E}$ and a point $\mathcal{E}$ in $\mathfrak{E}$ is of the form:

\begin{equation}
\label{ens}
\mathcal{E}= \{ p_i >0, \; \ketbra{\psi_i}{\psi_i} \}_{i=1}^{m} 
\end{equation}

Consider the space of pure state ensembles. When elements in $\mathcal{E}$ are indexed , $\mathcal{E}$ can correspond to it a gram matrix, G. This gram matrix should be $m \times m$, positive definite and have trace 1. Changing the index-sequence of elements in $\mathcal{E}$ is equivalent to performing a permuation transformation on the gram matrix. Additionally, by changing the phases associated with the pure states in $\mathcal{E}$ - $\ket{\widetilde{\psi}_j} \longrightarrow  e^{i \phi_j} \ket{\widetilde{\psi}_j} $, the gram matrix transforms under a diagonal unitary transformation. Varying these phases doesn't change the state $p_i \ketbra{\psi_i}{\psi_i}$ and doesn't change the ensemble. Yet the gram matrix changes. Thus, for the same ensemble, the corresponding gram matrix can vary upto a complex permutation. We want a more faithful representation so we define an equivalence class on the set of $m \times m$ positive definite matrices with trace 1\footnote{Two matrices from this set are equivalent iff 
one is a complex permutation of other other.}. We denote the quotient space by $\mathfrak{G}$. 
To represent elements in $\mathfrak{G}$ we adopt two conventions - (i) (ordering convention) $p_{i} \geq p_{i+1}$; in case of equality, $|G_{i \; i+1}| \geq  |G_{i+1 \; i+2}|$ and so on and so forth and (ii) (phase convention, after satisfying the ordering convention) $G_{i \; i+1} = G_{i+1 \; i} \; \forall 1\leq i \leq m$.

\begin{equation}
\label{GothicG}
\mathfrak{G} = \{G \text{ is m}\times \text{m} \;|\; G >0, \; \text{Tr}(G)=1; \; \text{G obeys ordering and phase conventions.} \}
\end{equation}

$\mathfrak{G}$ is a convex set. This fact will come in handy later on.

$\forall \; \mathcal{E} \, \in \, \mathfrak{E}$\footnote{Upto a unitary transformation of its elements: $\ketbra{\psi_i}{\psi_i} \longrightarrow U \ketbra{\psi_i}{\psi_i}U^\dag \; \forall \; 1 \leq i \leq m$}, $\exists$ a unique gram matrix, $\text{G} \in \mathfrak{G}$ corresponding to it. Let this mapping be represented as: $\mathcal{G}(\mathcal{E})  = G$.

Consider a trajectory in the space $\mathfrak{E}$:

\begin{equation}
\label{trE}
E: [ 0,1] \longrightarrow \mathfrak{E}
\end{equation}

Thus $E(t)$ represents a point in $\mathfrak{E}$ for $0 \leq t \leq 1$. The function E is well behaved with respect to the variable t\footnote{i.e. the a priori probabilities and matrix elements of all pure states in E(t) are analytic functions of t}. How does $P_{s}^{max}$ vary along this trajectory? From the definition of $P_s$, i.e. \eqref{Ps}, we see that it is continuous both in the ensemble input $\{p_i, \ketbra{\psi_i}{\psi_i} \}$ and the POVM input $\{ \Pi_i \}$. If $P_{s}^{max}$ were to vary discontinuously as t is varied from 0 to 1, the behaviour of $P_s$ would have to suffer discontinuities as a function of t too. Hence $P_{s}^{max}$ varies continuously as t varies from 0 to 1\footnote{This holds regardless of whether the ensemble is linearly independent or not.}.Corresponding to the map E(t) we get a similar map: $G:\; [0,1] \; \longrightarrow \mathfrak{G}$ where $G(t)\, =\, \mathcal{G}(E(t))$.

The the set of m-element POVMs forms a convex set. A POVM is generally an ordered set of m positive operators which sum to the $\mathbb{1}$. Convex sum of two POVMs is given by point-wise addition of ith elements: - $\{ V^{(1)}_{i} \}_{i=1}^{m}$ and $\{ V^{(2)}_{i} \}_{i=1}^{m}$, the ith element of the new POVM $\{ V_i \}_{i=1}^{m}$ is the convex sum of respective ith elements from $\{ V^{(1)}_{i} \}_{i=1}^{m}$ and $\{ V^{(2)}_{i} \}_{i=1}^{m}$ i.e. $V_i = p V^{(1)}_i + (1-p) V^{(2)}_i$.

Suppose $\{ V^{(1)}_{i} \}_{i=1}^{m}$ and $\{ V^{(2)}_{i} \}_{i=1}^{m}$ are projective measurments. Any convex combination of these projective measurments is a projective measurement iff either if (i) $p=0$ ( or $p=1$) OR (ii)  $\{ V^{(1)}_{i} \}_{i=1}^{m}=\{ V^{(2)}_{i} \}_{i=1}^{m}$\footnote{Both \textbf{ordered sets} have to be equal. Also, this holds even for projective measurements which aren't rank 1.}.

The optimal POVM for linearly independent pure states \emph{has to be} an m-element projective measurement. This implies that the optimal POVM \emph{has to be} unique.

Now as t varies from 0 to 1 how does the optimal POVM vary? The optimal POVM will be an m-element rank one projective measurement, so we can confine our focus on this space itself. The first thing to notice is that the uniqueness of the optimal POVM implies that we, actually, \emph{can} define a mapping $V : \;[0,1] \longrightarrow \mathfrak{V}$ such that $V(t)$ is the optimal POVM for MED of $E(t)$. Is $V(t)$ continuous in t\footnote{i.e. matrix elements of V(t) continous in t}? Suppose not; let there be a jump at some point $t_0$ so that $ \lim_{\substack{ \; t \underset{one-side}\longrightarrow \; t_0}} \; V(t)\; \neq \; V(t_0)$. Since $P_s$ is continuous function in both its ensemble input and its POVM input, we require that $ P_{s}^{max}(t_0) > \lim_{\substack{t \rightarrow t_0}} P_{s}^{max}(t)$: a contradiction. Thus $V(t)$ has to vary continuously as a function of t. 

From \eqref{B1} we know that V(t) has to be the pretty good measurement associated with some ensemble of the same states but different a priori probability. At the gram matix level, both ensembles are related by $G \longrightarrow \frac{1}{\text{Tr}(DGD)}DGD$ where $D$ is a positive diagonal matrix such that the positive $ \sqrt{DGD}$ (which is positive definite) has diagonal elements $a_{ii}^{2}$ (see \eqref{SSGMM1} ).

We can construct V(t) from $\sqrt{\text{D(t) G(t) D(t)}}$ using simple arithmetic operations of the matrix elements of the latter(see \eqref{SSGMM}). Thus, if D(t) and D(t) G(t)D(t) vary analytically as a function of t, then so does V(t). To inspect if the former is true, we employ the implicit function theorem. 

Implicit Function Theorem:  Let $ \{ y_{i} \}_{i=1}^{N}$ be N functions (real or complex) of the independent variables - $\{ t, f_1, f_2, \cdots, f_N \}$ where the variables $\{ f_i \}_i$, which are N in number too. Let $(\tau, \phi_1, \phi_3, \cdots, \phi_N )$ be a point such that $y_i(\tau, \phi_1, \phi_2, \cdots, \phi_N )=0 \; \forall \; 1 \leq i \leq N$. If the Jacobian matrix $ J_{i,j} = \frac{\partial y_i}{\partial f_j}$ is invertible at $(\tau, \phi_1, \phi_3, \cdots, \phi_N )$ then there exists some open neighbourhood, T, containing $\tau$:  for which there exist open neighbourhoods $S_i$ containing $\phi_i$ such that $f_i: T \longrightarrow S_i$ can be defined, so that $y_i(t, f_1(t), f_2(t), \cdots, f_N(t) )=0 \; \forall \; 1 \leq i \leq N, \quad \forall \; t \; \in \; T$. That is

\begin{equation}
\label{ImpDef}
\{ (t,\vec{f}) \in T \times S \quad | \quad\vec{y}(t,\vec{f}) \; = \;0\;  \} = \{\; (t,\vec{f(t)}) \quad |\quad t\;\in \; T \text{ and } \; y(t,\vec{f(t)}) \; = \;0 \}
\end{equation} where $S = S_1 \times S_2 \times \cdots \times S_N$.

Simply put, the implict function theorem gives sufficient conditions for the variables $f_i$ to implicitly depend on the variable t near some point $(\tau, \phi_i)$ so that  in some neighbourhood of this point $\vec{y} \; (t, \; \vec{f(t)} )$ is constant in t. 

Analytic Implicit Function Theorem: Furthermore if $y_i$ is an analytic function in the variables $ f_i$ then the implicit dependence of $ f_i $ on t will be analytic too.

From \eqref{B1} we get that:

\begin{equation}
\label{DGD1}
(D G^{\frac{1}{2}} \widetilde{U})^2 - DGD =0
\end{equation}

where $D G^{\frac{1}{2}} \widetilde{U}$ is the positive square root of DGD and $(D G^{\frac{1}{2}} \widetilde{U})_{ii}=a_{ii}^{2}$. In our application of IFT, t is the independent variable. The implicit variables $f_{ij}$ and $a_i$ are defined as:

\begin{align}
\label{1implicit1}
f_{ij}(t) \; \longrightarrow \; (D G^{\frac{1}{2}} \widetilde{U})_{ij}  \; \text{when }i \neq j \\
\label{1implicit2}
 a_{i}(t) \; \longrightarrow \; \sqrt{ (D G^{\frac{1}{2}} \widetilde{U})_{ii} } 
\end{align}

The implict variables -$\{ a_{i}, f_{ij} | i \neq j \}$  are $m^2$ in number. Here these variables are complex, thus $f_{ij}$ and $f_{ji}$ are explicitly taken different.

Define F(t) to be an $m \times m$ matrix defined by:

\begin{align}
\label{F1}
F_{ij}(t)= f_{ij}(t)\; \text{when }i \neq j \\
\label{F2}
F_{ii}(t)= {a_i (t) }^2
\end{align}

Consider the matrix equation:

\begin{equation}
\label{EF}
(F(t))^2-D(t)G(t)D(t)= 0
\end{equation}

This the same as \eqref{DGD1}.

Define a new matrix Y(t) as:\begin{equation}
\label{J1}
y_{ij}(t) = ((F(t))^2-D(t)G(t)D(t))_{ij}
\end{equation} where $(Y(t))_{ij} = y_{ij}(t)$.

Hence, we want to the functions $\{ a_i(t), f_{ij}(t) \}$ to vary implicitly on t in a manner such that the functions $y_{ij}(t)=0$. 

The Jacobian is given by \begin{align}
\label{Jacob1}
& J_{(ij),(kl)} = \frac{\partial y_{ij} }{\partial f_{kl} } \; \text{ where } \; k \neq l \\
\label{Jacob2}
& J_{(ij),(kk)} = \frac{\partial y_{ij} }{\partial a_{k} }
\end{align}
The Jacobian matrix is an $m^2 \times m^2$ matrix. Now a matrix is invertible iff it's rows (or columns) are linearly independent. We de-vectorize the rows of the Jacobian matrix and bring them in the form of $m \times m$ matrices. If \emph{these} matrices (which are $m^2$ in number) are linearly independent then that proves that the Jacobian is invertible.

Define matrix $M^{i,j}$ by:

\begin{equation}
\label{devec}
M^{ij}_{kl}= J_{(ij),(kl)}
\end{equation}

This matrix is of the form:

\begin{equation}
\label{Mij}
M_{ij} =\bordermatrix { ~     &          &         &        & \underset{\downarrow}{(\text{j}^{\text{th}}\text{ column})} &  &  \cr
&  0       & 0       & \cdots & X				   & \cdots & 0 \cr
&  0       & 0       & \cdots & X 				   & \cdots & 0 \cr
&  \vdots  & \vdots  & \vdots & X 				   & \vdots & 0 \cr
(\text{i}^{\text{th}}\text{ row} \rightarrow) &  X       & X       & X      & X 				   & X      & X \cr
&  0       & 0       & \cdots & X 				   & \cdots & 0 \cr
&  \vdots  & \vdots  & \vdots & X 				   & \ddots & 0 \cr
&  0       & 0       & \cdots & X 				   & \cdots & 0 \cr}\end{equation}

where the X's are polynomials in the implicit variables $\{ a_i(t), f_{ij}(t)\}$ and matrix elements of $G(t)$. While it is reasonable to expect that the matrices $M^{ij}$ are linearly independent (implying the the Jacobian is invertible) unless we know how the implicit variables behave as functions of t, we can't conclude anything about their linear independence for all points in $\mathfrak{G}$. This puts us in a cyclic conundrum. The implicit function theorem provides a sufficient but not necessary condition for the existence of such implicit functions. That is, there are pathological examples where an implicit function exists even at points where Jacobian is singular. In this case, our inability to establish that the Jacobian isn't singular at all points $G \in \mathfrak{G}$ isn't a cause for worry. This is because we know beforehand \cite{Carlos}\cite{Bela} that such an implicit function exists, that it is continuous and furthermore we know that it is globally one-to-one with respect G $\in \mathfrak{G}$.
 
Implicit dependence does not exist when a critical point is a bifurcation point. In our case we know that such a point cannot exist. Otherwise, we'd have two different optimal POVM for the same point in $\mathfrak{G}$. And we know that this cannot happen for linearly independent ensembles. 

Now that we have dealt with the question of the invertibility of the Jacobian we note that the $y_{ij}$ are analytic functions of the variables $t; a_i, \;f_{ij}$. Thus $a_i(t), \; f_{ij}(t)$ have to be analytic in terms of t\footnote{From the analytic implicit function theorem.}. This piece of information is important because it tells us that the diagonal matrix D(t) varies analytically with t. And this, in turn implies, as we argued before, that the optimal POVM will vary analytically with the ensemble V(t). 

\subsection{Analytic Continuation of the Optimal POVM}

Having established that the optimal POVM will vary analytically as a function of the ensemble at all points $G \, \in \, \mathfrak{G}$, we ask the following question: Can one \textbf{drag} the optimal POVM from one ensemble to another by employing some technique? 

For this, first of all, we need a trajectory. Using the mapping $\mathcal{G}$, one can find a simple trajectory between two ensembles. Let$E_1$ represent the first ensemble and $G_1 = \mathcal{G}(E_1)$. Let the optimum POVM for $E_1$ be known. That is to say, we know what $D_1$ is. Let $E_2$ be another ensemble and $G_2 = \mathcal{G}(E_2)$. We want to obtain $D_2$. Since $\mathfrak{G}$ is a convex set, one can obtain a simple linear trajectory between them:\begin{equation}
\label{trajectory}
G(t) = (1-t)\; G_1 + t \; G_2 \: \: \text{where } t \in \; [0,1]
\end{equation} For any value of $ t \in \; [0,1]$, $G(t) \in \mathfrak{G}$.

We've established the trajectory.Next we need some differential equations which we use to drag D(t) from $D_1$ to $D_2$. For this employ IFT again: take the total derivative w.r.t of the equation \eqref{J1} for all matrix elements and set it equal to zero.\begin{equation}
\label{y0}
\frac{d \; y_{kl}}{d \; t} =\zeta_{kl}(t; a_i(t), f_{ij}(t), \frac{d a_i(t)}{dt}, \frac{d f_{ij}(t)}{dt})   = 0 \: \: \forall \: 1 \; \leq k,l \; \leq m
\end{equation}This gives us a set of $m^2$ first order coupled ordinary differential equations for the implicit variables $a_i(t), \; f_{ij}(t)$. 

The form of $\zeta_{kl}$  in terms of $a_i, f_{ij}, \pd{a_i}, \pd{f_{ij}}$ is rather complicated and depends on the value of m. As an illustration we write down what $\zeta_{kl}$ look like for m = 2:    
\begin{eqnarray}
\label{m2p11}
\zeta_{11} &  = &  4 a_{1}(t)^3 \pd{a_1(t)}+ f_{12}(t) \pd{f_{21}(t)}+f_{21}(t) \pd{f_{12}(t)}- 2 a_1(t) g_{11}(t)\pd{a_1(t)}-{a_{1}(t)}^2\pd{g_{11}(t)} \\ & = & 0 \notag \\
\label{m2p12}
\zeta_{12} &  = &   (a_1(t)^2 + a_2(t)^2)\pd{f_{12}(t)}+(2a_1(t)f_{12}(t)-a_2(t)g_{12}(t))\pd{a_1(t)} +\\ &  &  (2a_2(t)f_{12}(t)-a_1(t)g_{12}(t))\pd{a_2(t)} -a_1(t)a_2(t) \pd{g_{12}(t)} \notag  \\ &  = &0 \notag \\
\label{m2p21}
\zeta_{21} & = &  (a_1(t)^2 + a_2(t)^2)\pd{f_{21}(t)}+(2a_1(t)f_{21}(t)-a_2(t)g_{21}(t))\pd{a_1(t)}  \\ & & +  (2a_2(t)f_{21}(t)-a_1(t)g_{21}(t))\pd{a_2(t)} -a_1(t)a_2(t) \pd{g_{21}(t)} \notag  \\ & = &0 \notag \\
\label{m2p22}
\zeta_{22}  & =  &  4 a_{2}(t)^3 \pd{a_2(t)}+ f_{12}(t) \pd{f_{21}(t)}+f_{21}(t) \pd{f_{12}(t)}- 2 a_2(t) g_{22}(t)\pd{a_2(t)}-{a_{2}(t)}^2\pd{g_{22}(t)} \\ &  = & 0 \notag
\end{eqnarray}

Here $ g_{ij}(t) = {G(t)}_{ij}$ is where the explicit time dependence comes into the equations.

These equations become more complicated as m increases. These equations are hermiticity preserving i.e. $a_i(t) \in \mathbb{R} \Rightarrow a_i(t+\delta)\, \in \, \mathbb{R}$ and and $f_{ij}(t) = f_{ji}(t)^* \Rightarrow f_{ij}(t+\delta) = f_{ji}(t+\delta)^*$ for infinitesimal $\delta$.

The equations give us a way of obtaining the optimal POVM by dragging the solution from a point where the solution is known to our desired point where the solution is unknown. It is worth mentioning how everything fits into the technique. We start from a point whose solution we know and  we drag the solution from one point to another using our trajectory. This dragging is based on the rule that \eqref{StG1} (and hence \eqref{St}) is always satisfied, i.e. if at a point $t$ we have $(a_i, \, f_{ij} ) \; \ni$ \eqref{StG1} is satisfied we ask what values for $(a_i, f_{ij})$ will satisfy \eqref{StG1} at the point $t + \delta$. Note that \eqref{StG1}, by itself, is merely a necessary condition. \eqref{Glb} needs to also be satisfied for the required solution. Now \eqref{Glb} $\Longleftrightarrow \, F(t)>0$. How do we know that given a trajectory within $\mathfrak{G}$ the differential equations will preserve the positivity of the matrix F(t)? F(0) is positive definite. Let's suppose there is a point, $t_1$ where 
one of the eigenvalues of F($t_1$) is negative. That means that there must have been some previous point (say, $t_0$ at which this particular eigenvalue would have been 0. From \eqref{EF} its implied that the matrix D($t_0$) must also be singular at this point. From t=0 to t=$t_0$, $F(t)>0$ and $D(t)>0$. This implies that for all ensembles from t=0 to t=$t_0$, solution has been obtained. Examine that as we approach the limit $t \underset{\text{from 0}}{\longrightarrow} t_0$, one of the values of $a_i \longrightarrow 0$. Such scenarios can only arise when (i) $p_i(t) \longrightarrow 0$. Then at $t=t_0$, we just have a set of m-1 states in our ensemble. OR (ii) The part of  $\ket{\psi_i(t)}$ that is perpendiculr to other states in the ensembles tends to zero, i.e. $\underset{(t-t_0)^{-}}{\text{limit }}\ket{u_i(t)}= 0$. In either case $G(t_0)\, \notin\, \mathfrak{G}$ which implies that G(t) isn't continuous in $\mathfrak{G}$. Hence we can conclude that as long as the trajectory is safely within $\mathfrak{G}$, 
we won't encounter this problem.

We employed this method to obtain solutions for various linearly independent pure state ensembles for cases: m = 2, 3, 4 and 5. In all the different cases our starting point was the equiprobable orthogonal ensemble for which the solution is trivial i.e. $G_1 = \frac{1}{m} \, \mathbb{1}$. We employed the RK4 method to solve for many unknown ensembles (generated randomly). And in all cases we managed to construct a POVM which satisfied \eqref{St} and \eqref{Glb} i.e. we construted the optimal POVM for this case. 

One can obtain the solution with as much precision as desired using RK4. For RK4, the local truncation varies as $\mathcal{O}(h^5)$ and the total accumulated error as $\mathcal{O}(h^4)$. Our method gives us a way of measuring the error without having to compare results with other techniques \cite{Jezek}\cite{Hel2}\cite{Eldar}. Consider \eqref{EF}. The deviation of the RHS from 0 is an indication of how much error has seeped in. Thus we plot Hilbert-Schmidt norm of the RHS of \eqref{EF} as a function of the iteration to get an idea of how much error accumulates. We illustrate with an example when m = 5. Let the gram matrix of the ensemble be given by:

\[\arraycolsep=1.4pt\def\arraystretch{2.2}
G  =  \begin{pmatrix}
     & 0.3 & \sqrt{0.06}(0.2 + i 0.1) & \sqrt{0.06}(0.1) & \sqrt{0.045}(0.1) & \sqrt{0.045}(0.1)  \\
     & \sqrt{0.06}(0.2 - i 0.1) & 0.2 & 0.06 & \sqrt{0.03}(0.2+i 0.2) & \sqrt{0.03}(0.1)  \\
     & \sqrt{0.06}(0.1) & 0.06 & 0.2 & \sqrt{0.03}(0.2+i 0.05) & \sqrt{0.03}(0.3 +i0.2) \\
     & \sqrt{0.045}(0.1) & (0.2-i 0.2)\sqrt{0.03} & \sqrt{0.03}(0.1-i0.05) & 0.15  &  (0.15)(0.2 +i0.3)  \\
     & \sqrt{0.045}(0.1) & (0.1)\sqrt{0.03} &(0.3-i0.3) \sqrt{0.03} &(0.15)(0.2 -i0.3)   &  0.15   
    \end{pmatrix}
\]

The interval length is given by $\text{h }= 10^{-3}$ and the number of iterations is 1000. We plot the log of the Hilbert Schmidt norm of the RHS of \eqref{EF}. 

\begin{figure}
    \centering
    \includegraphics[width=440pt]{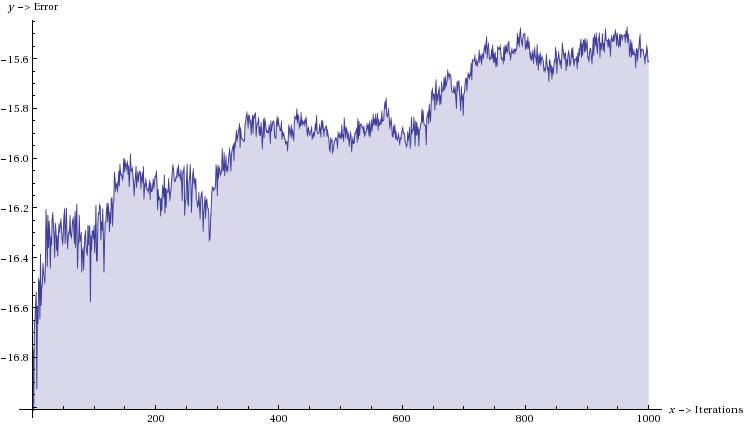}
    \begin{minipage}{440pt}
    \footnotesize
    \emph{y- axis :Log of HS norm of RHS of \eqref{EF}, x-axis: No. of iterations. One can see the gradual increase the norm from $~-16.8$ when $ 1 \leq x \leq 10$ to $-15.7$ for $ 980 \leq x \leq 1000$. This shows that the truncation error is ~ $10^{-16}$ and the total accumulated error is ~ $10^{-15}$ which shows a pretty good performance for RK4.}
    \end{minipage}
    \caption{Error vs Iteration No.}
    \label{erit}
    \end{figure}
From the figure we see that for the first few iterations the truncation error is of order ~ $-16.8$ and over the last few iterations total accumulated error is of order ~ $-15.6$ which is well within the error-performance given by RK4. For many randomly selected ensembles the local truncation error is of order ~ $-16$ and total accumulated error is of order $-15$ - again well within the RK4 error margin. This shows that this method is a reliable method to obtain the optimal POVM for linearly independent pure states. 

Where does the method? If the ensemble is nearly linearly dependent (i.e.  $\exists \, i \, \ni \, | \text{either} \braket{\psi_i}{u_i}| $ is \emph{very small} OR $p_i$ is \emph{very small}\footnote{\emph{very small}: the order of the local truncation error}) then the method is likely to fail. 

\section{Linearly Dependent States}

This method generally won't work when for linearly dependent ensembles. Three major problems are identified:

\textbf{IA. Why any such technique won't work generally: Optimal POVM doesn't always vary smoothly}

From \cite{Carlos}\cite{Bela} we see that for an ensemble of linearly dependent pure states $\{\widetilde{p}_i \geq 0, \ketbra{\psi_i}{\psi_i} \}_{i=1}^{m}$, the ensemble for which the PGM of the former will discriminate optimally is given by $\{ p_i > 0, \ketbra{\psi_i}{\psi_i}\}_{i=1}^{m}$ satisfying the rule:

\begin{align*}
\label{ldrule}
 p_i & \leq \frac{c}{    \bra{\psi_i} \sqrt{   \widetilde{\rho}^{-1}   } \ket{\psi_i} } \text{ when } \widetilde{p}_i =0 \\
 p_i & = \frac{c}{\bra{\psi_i} \sqrt{\widetilde{\rho}^{-1}} \ket{\psi_i}} \text{ when } \widetilde{p}_i >0 
\end{align*}

where c is some normalization constant and $\sqrt{   \widetilde{\rho}^{-1}   }$ is the inverse of the positive square root for the ensemble $\{ \widetilde{p}_i, \psi_i\}$.

Our problem is opposite: we need to find the optimal POVM given the ensemble. But from the above theorem one can conclude that as one varies the ensemble of states, the optimal POVM will not necessary vary in a smooth fashion. Also, the above mapping isn't one-to-one; i.e. for the same ensemble one can find two or more ensembles whose PGM's will optimally discriminate among the members of the former ensemble. 

\textbf{IB. Why any such technique won't work generally: Optimal POVM can vary discontinuously}

For linearly dependent pure states, the optimal POVM for an ensemble need not be unique. For example the optimal POVM for the equiprobable ensemble $\{ \frac{1}{4}\ketbra{0}{0},\: \: \frac{1}{4}\ketbra{1}{1}, \: \: \frac{1}{4}\ketbra{+}{+}, \: \: \frac{1}{4}\ketbra{-}{-} \}$ \footnote{ $\ket{\pm} = \frac{1}{\sqrt{2}}( \ket{0} \pm \bra{1})$} has two extremal optimal solutions - $\{\ketbra{0}{0}, \: \: \ketbra{1}{1}    \}$ and $\{\ketbra{+}{+}, \: \: \ketbra{-}{-} \}$ and hence any convex combination of these two is also an optimal solution. Were this ensemble to lie on a trajectory, the optimal solution of points on one side of this ensemble could be closer to $\{\ketbra{0}{0}, \: \: \ketbra{1}{1}    \}$ than $\{\ketbra{+}{+}, \: \: \ketbra{-}{-} \}$ and vice versa for points on the other side of this ensemble. This shows that the optimal POVM can vary discontinuously. 

\textbf{II. Why this technique won't work: $\nexists$ a set of states $\{ \ket{u_i} \; | \; \braket{   \widetilde{\psi}_i   }{  \widetilde{u}_j   } = \delta_{i,j} \}$ when $\{ \ket{\psi_i} \}_i$ are linearly dependent}

Our technique relies upon casting the problem in a representation that exists only when the states are linearly independent. The speciality with this representation is that the one can use the implicit function theorem which demands that the number of function $y_i$ and the number of implicit variables $f_i$ have to be equal.

\end{document}